\documentclass{ws-ijmpa}
\usepackage[super,compress]{cite}
\usepackage{graphicx}
\begin{document}
\markboth{Eric Sharpe}{A survey of recent developments in GLSMs}

%
\catchline{}{}{}{}{}
%

\title{A survey of some recent developments in GLSMs}

\author{Eric Sharpe}
\address{Virginia Tech\\
Department of Physics\\
850 West Campus Drive\\
Blacksburg, VA 24061\\
ersharpe@vt.edu}

\maketitle

\begin{history}
\received{Day Month Year}
\revised{Day Month Year}
\end{history}

\begin{abstract}
In this article we briefly survey some developments in
gauged linear sigma models (GLSMs).  Specifically, we give an overview of
progress on constructions of GLSMs for various geometries,
GLSM-based computations of quantum cohomology, 
quantum sheaf cohomology, and
quantum K theory rings, 
as well as two-dimensional abelian
and nonabelian mirror constructions.
(Contribution to the proceedings of {\it Gauged Linear Sigma Models@30} 
(Simons Center,
Stony Brook, May 2023).)

\end{abstract}

\ccode{PACS numbers: 11.25.Mj, 11.10.Kk}

\tableofcontents

\section{Introduction}

Gauged linear sigma models (GLSMs) were first described
thirty years ago \cite{Witten:1993yc}.  They quickly became vital
tools in string compactifications, still used and developed today.
The goal of this article (and the corresponding talk at the workshop
{\it GLSMs@30})
is to briefly survey
some of the developments and current research areas in GLSMs.  
To be clear, there is not
enough space to describe, much less give justice to, everything
that has been developed or is being researched,
but we do hope to outline many areas, and will reference related talks
that took place at {\it GLSMs@30}.

\section{Constructions of geometries}

Originally, GLSMs were used to give physical realizations 
of geometries of the form of complete
intersections in symplectic
quotients ${\mathbb C}^n // G$.
Briefly, the idea is to realize ${\mathbb C}^n // G$ as a two-dimensional
supersymmetric $G$-gauge theory with matter fields corresponding
to ${\mathbb C}^n$, plus additional matter and a 
superpotential for which
the complete intersection is the critical locus.

For example, to describe a hypersurface $\{ G = 0 \} \subset
{\mathbb C}^n // G$, one starts with
a gauge theory describing ${\mathbb C}^n // G$, and adds a chiral superfield
$p$ and a superpotential
$W = p G$, where $p$ is chosen to transform under the action of
$G$ in such a way that $W$ is gauge-invariant.
If the hypersurface is smooth, then the critical locus reduces to
\begin{equation}
\{ p = 0 \} \cap \{ G = 0 \},
\end{equation}
which is the desired hypersurface in ${\mathbb C}^n//G$.
We will refer to this as a ``perturbative'' description.

Nowadays we know of two alternative 
mechanisms that can be used to realize
geometries:
\begin{itemize}
\item Strong coupling effects in two-dimensional gauge theories can
restrict the space of vacua.  The prototype for this is the
GLSM for the Grassmannian-Pfaffian system 
\cite{Hori:2006dk}.
\item Decomposition \cite{Hellerman:2006zs,Sharpe:2022ene} 
locally realizes a branched cover.
Prototypes for this are GLSMs relating complete intersections of
quadrics to branched covers
\cite{Caldararu:2010ljp}.
\end{itemize}

Let's quickly walk through each of these in turn.

First, we consider nonperturbative constructions of Pfaffians
\cite{Hori:2006dk}.
The prototypical example is the GLSM for the complete intersection
of seven hyperplanes in the Grassmannian $G(2,7)$, which
is denoted $G(2,7)[1^7]$.  
This GLSM is a $U(2)$ gauge theory with 7 fundamentals $\phi_i^a$
plus $7$ chiral superfields denoted $p_{\alpha}$ which
are charged under $\det U(2)$, with a superpotential
\begin{equation}
W \: = \: \sum_{\alpha} p_{\alpha} G_{\alpha}\left( \epsilon_{ab}
\phi_i^a \phi_j^b \right) \: = \:
\sum_{ij} \epsilon_{ab} \phi^a_i \phi^b_j A^{ij}(p).
\end{equation}
For $r \gg 0$, this GLSM describes $G(2,7)[1^7]$, by the usual analysis.
For $r \ll 0$, the analysis of this GLSM utilizes results from the
strongly-coupled gauge theory.  Working locally in a Born-Oppenheimer
approximation along the space of vevs of the $p_{\alpha}$ fields,
\begin{itemize}
\item loci with one massless doublet (generic case) have no susy vacua,
\item loci with three massless doublets have one susy vacuum.
\end{itemize}
The resulting theory, the loci with 3 massless doublets,
describe a Pfaffian variety inside the projective space
${\mathbb P}^6$ defined by the $p_{\alpha}$.

Next, we turn to nonperturbative constructions of branched covers
\cite{Caldararu:2010ljp}.
A simple example involves the GLSM for ${\mathbb P}^3[2,2]$.
This is a $U(1)$ gauge theory, with four chiral multiplets $\phi_i$ of
charge $+1$, two chiral multiplets $p_{\alpha}$ of charge $-2$,
and a superpotential
\begin{equation}
W \: = \: \sum_{\alpha} p_{\alpha} G_{\alpha}(\phi) \: = \:
\sum_{ij} S^{ij}(p) \phi_i \phi_j.
\end{equation}
For $r \gg 0$, this describes ${\mathbb P}^3[2,2] = T^2$, by the usual
analysis.
For $r \ll 0$, working locally in a Born-Oppenheimer approximation on the
space of vevs of the $p_{\alpha}$ fields, which is ${\mathbb P}^1$,
the $S^{ij}$ acts as a mass
matrix for the charge $1$ fields $\phi_i$.  
To correctly analyze this phase, we must use the fact that
at low energies, the gauge theory (generically) has a trivially-acting
${\mathbb Z}_2 \subset U(1)$, hence a ${\mathbb Z}_2$ one-form
symmetry, and so by decomposition \cite{Hellerman:2006zs,Sharpe:2022ene}, 
is (generically) a double cover, away from the locus $\{ \det S = 0 \}$,
where some of the $\phi_i$ become massless.
The resulting geometry is a double cover
of ${\mathbb P}^1$ (the space of vevs of the $p_{\alpha}$), branched
over a degree-four locus ($\{ \det S = 0 \}$), which is another $T^2$.

The GLSM for ${\mathbb P}^5[2,2,2] = K3$ can be analyzed very similarly.
The $r \ll 0$ phase is a branched double cover of ${\mathbb P}^2$,
branched over a degree 6 locus, which is another $K3$.

Starting in 3-folds, these examples becomes more interesting.
The GLSM for ${\mathbb P}^7[2,2,2,2]$ describes a noncommutative resolution
of a branched double cover, defined  
\cite{kuz1,kuz2,kuz3}in terms of derived categories.
In particlar, the GLSM gives a UV representation of a closed string CFT
for a noncommutative resolution.  The noncommutative structure is detected
physically by studying matrix factorizations in (hybrid) Landau-Ginzburg
phases -- in other words, by examining D-branes.

These noncommutative resolutions were discussed elsewhere at this meeting,
in talks of S.~Katz, T.~Schimannek, M.~Romo, and J.~Guo.

Another property of these 3-fold examples (both the Grassmannian/Pfaffian
and the branched covers) is that the different GLSM phases are not birational
to one another.
This contradicted folklore of the time, which said that all (geometric)
phases of a single GLSM
should be birational.
Instead, these phases are related by homological projective duality
\cite{kuz1,kuz2,kuz3}.
This has been studied in this context in
mathematics, in variations of GIT quotients, see
for example \cite{bfk,bdfik,hl1,hl2,r1,r2,r3}.
Homological projective duality is beyond the scope of this overview,
but was discussed elsewhere at this meeting, in talks of
J.~Guo and M.~Romo.

Nowadays, we can also realize similar effects perturbatively.
For example, Pfaffians can be described via the PAX and PAXY models 
\cite{Jockers:2012zr}.
Perturbative and nonperturbative constructions can be exchanged by
dualities, see e.g.~\cite{Caldararu:2017usq}.

\section{Quantum cohomology and 2d mirrors}

One of the original applications of GLSMs was to make predictions for
quantum cohomology rings of Fano toric varieties.  For such spaces, 
we can use the GLSM to replace counting rational curves with an
algebraic computation, on the Coulomb branch, that encodes the
same result.  In particular, quantum 
cohomology can be seen in a Coulomb branch computation.
For example, under RG flow, the GLSM for ${\mathbb P}^n$ describes a space that shrinks
to (classical) zero size, and then onto the Coulomb branch, where
quantum cohomology is describe as the classical critical locus of a 
twisted one-loop effective superpotential, instead of as a sum
over rational curves.

For Fano symplectic quotients ${\mathbb C}^n//G$ 
for $G =  U(1)^k$, the 
twisted one-loop effective superpotential is
of the form \cite{Morrison:1994fr}
\begin{equation}
\tilde{W}(\sigma) \: = \: 
\sum_{a=1}^k \sigma_a \left[ \tau_a \: + \:
\sum_i Q_i^a \left( \ln \left( \sum_{b=1}^k Q_i^b \sigma_b \right) - 1
\right) \right],
\end{equation}
and the resulting critical locus $\{ \partial \tilde{W} / \partial \sigma_a
= 0 \}$ is given by \cite{Morrison:1994fr}
\begin{equation}
\prod_i \left( \sum_b Q_i^b \sigma_b \right)^{Q_i^a} \: = \:
\exp\left( 2 \pi i \tau_a \right) \: = \: q_a.
\end{equation}
If the theory in the IR is a pure Coulomb branch, then these are the
quantum cohomology relations.

To make this more concrete, let us specialize to ${\mathbb P}^n$.
Under RG flow, the GLSM for ${\mathbb P}^n$ describes a space that shrinks to
(classical) zero size, and then onto the Coulomb branch.  The one-loop
twisted effective superpotential is
\begin{equation}
\tilde{W} \: = \: \sigma \left[ \tau + \sum_{i=1}^{n+1} \left( \ln \sigma -
1 \right) \right],
\end{equation}
which has critical locus given by the solution to
\begin{equation}
\frac{\partial \tilde{W}}{\partial \sigma} \: = \:
\tau + \ln \left( \sigma^{n+1} \right) \: = \: 0,
\end{equation}
namely
\begin{equation}
\sigma^{n+1} \: = \: \exp(-\tau) \: = \: q.
\end{equation}
This is precisely the well-known quantum cohomology ring relation for
${\mathbb P}^n$, 
identifying $\sigma$ with a generator of $H^2( {\mathbb P}^n )$.

The same ideas also apply to nonabelian GLSMs, meaning,
GLSMs describing spaces of the form ${\mathbb C}^n//G$ for nonabelian $G$
(and subvarieties thereof).  For Fano ${\mathbb C}^n//G$, RG flow
again drives the GLSM out of a geometric phase and onto the Coulomb branch.
Again the quantum cohomology ring arises as the critical locus of a 
superpotential, albeit with two subtleties:
\begin{itemize}
\item The Coulomb branch is a Weyl-group orbifold of the $\sigma$'s,
\item The Coulomb branch is 
an open subset of the space of $\sigma$'s -- an `excluded locus' is removed.
\end{itemize}

To make this discussion concrete, we turn to the example of the
Grassmannian $G(k,n)$ of $k$-planes in ${\mathbb C}^n$.
This can be described as the symplectic quotient ${\mathbb C}^{kn}//U(k)$,
where $U(k)$ acts as $n$ copies of the fundamental representation.
Here, the twisted one-loop effective superpotential is
\begin{eqnarray}
\tilde{W} & = & \sum_{a=1}^k \sigma_a \left[
- \ln \left( (-)^{k-1} q \right) \: + \:
\sum_{i,b} Q_{ib}^a \left( \ln \left( \sum_{c=1}^k Q^c_{ib} \sigma_c \right)
- 1 \right) \right],
\\
& = & \sum_{a=1}^k \sigma_a \left[
- \ln \left( (-)^{k-1} q \right) \: + \:
\sum_{i=1}^n \left( \ln \sigma_a - 1 \right) \right],
\end{eqnarray}
using the fact that $Q_{ia}^b = \delta^a_b$ for copies of the fundamental
representation.  In principle, the space of $\sigma$'s is orbifolded
by the Weyl group of $U(k)$ (namely, the symmetric group $S_k$), 
which acts by interchanging the $\sigma_a$,
and we also remove the `excluded locus' $\{ \sigma_a = \sigma_b, a \neq b \}$.
The critical locus is computed from
\begin{equation}
\frac{\partial \tilde{W}}{\partial \sigma_a} \: = \:
- \ln \left( (-)^{k-1} q \right) \: + \: \ln (\sigma_a)^n \: = \: 0,
\end{equation}
which implies
\begin{equation}  \label{eq:gkn:critlocus}
\left( \sigma_a \right)^n \: = \:  (-)^{k-1} q.
\end{equation}
It may not yet be manifest, but this defines the quantum cohomology
ring relation for $G(k,n)$.

As a quick consistency check, we compute the number of vacua.
The relation above is an order $n$ polynomial, so for each value of $a$,
there are $k$ solutions, hence $kn$ possible values altogether.
Taking into account the $S_k$ orbifold and the excluded locus, the number
of admissible solutions to the critical locus equation is
\begin{equation}
\left( \begin{array}{c}
n \\ k \end{array} \right) \: = \: \chi\left( G(k,n) \right),
\end{equation}
as expected.

To make the relation to the quantum cohomology ring of the Grassmannian
more clear, we can rewrite the critical locus equation~(\ref{eq:gkn:critlocus})
as follows.  First, note that the $\sigma_a$ are $k$ distinct roots of the
$n$th order polynomial
\begin{equation}
\xi^n \: + \: (-)^k q \: = \: 0.
\end{equation}
Let $\overline{\sigma}_{a'}$ denote the remaining $n-k$ roots.
From Vieta's theorem in algebra, the elementary symmetric polynomials
$e_{i}$ in the $\sigma_a$ and $\overline{\sigma}_{a'}$ obey
\begin{equation}
\sum_{r=0}^{n-k} e_{\ell-r}(\sigma) \, e_r(\overline{\sigma}) \: = \:
(-)^{n-k} q \, \delta_{\ell,n} \: + \: \delta_{\ell,0}.
\end{equation}
Define
\begin{equation}
c_t(\sigma) \: = \: \sum_{\ell=0}^k t^{\ell} e_{\ell}(\sigma)
\end{equation}
and similarly for $\overline{\sigma}$, and then the result above from
Vieta's theorem can be written
\begin{equation}
c_t(\sigma) \, c_t(\overline{\sigma}) \: = \: 1 \: + \: (-)^{n-k} q t^n,
\end{equation}
which is a standard expression for the quantum cohomology ring of
$G(k,n)$, see e.g.~\cite[equ'n (3.16)]{Witten:1993xi},
where we interpret $c_t(\sigma)$ as the total Chern class of the universal
subbundle $S$ on $G(k,n)$, and $c_t(\overline{\sigma})$ as the total
Chern class of the universal quotient bundle $Q$.

So far we have reviewed Coulomb-branch-based quantum cohomology computations
in GLSMs.  Another approach to these and related questions is to use
mirror symmetry, which we will review next.

First, we will quickly review abelian mirrors
\cite{Hori:2000kt,Morrison:1995yh}.  
Briefly, start with a $U(1)^r$ gauge theory with matter multiplets of
charges $\rho_i^a$, corresponding to a quotient
${\mathbb C}^n // U(1)^r$.
The mirror is a Landa-Ginzburg model, defined by the chiral superfields
\begin{itemize}
\item $\sigma_a$, $a \in \{1, \cdots, r\}$, $\sigma_a = \overline{D}_+ D_-
V_a$,
\item $Y^i$, mirror to the matter fields of the original theory,
with periodicities $Y^i \sim Y^i + 2 \pi i$,
\end{itemize}
with superpotential
\begin{equation}
W \: = \: \sum_{a=1}^r \sigma_a \left( \sum_i \rho_i^a Y^i - t_a \right)
\: + \: \sum_i \exp\left(-Y^i \right).
\end{equation}

Next, we turn to mirrors to ${\mathbb C}^n//G$ for $G$ nonabelian
\cite{Gu:2018fpm}.
Here, we pick a Cartan torus $U(1)^r \subseteq G$, $r$ the rank of $G$,
and let $\rho$
defining the representation of $G$ under which the matter multiplets
transform.  The mirror is a Weyl-group-orbifold of the Landau-Ginzburg
model defined by the fields
\begin{itemize}
\item $\sigma_a$, $a \in \{1, \cdots, r\}$, $\sigma_a = \overline{D}_+ D_-
V_a$, 
\item $Y^i$, mirror to the matter fields of the original theory,
\item $X_{\tilde{\mu}}$, in one-to-one correspondence with the
nonzero roots of ${\mathfrak g}$,
\end{itemize}
and superpotential
\begin{equation} \label{eq:W:nonabelian:mirror}
W \: = \: \sum_{a=1}^r \sigma_a \left( \sum_i \rho_i^a Y^i \: - \:
\sum_{\tilde{\mu}} \alpha^a_{\tilde{\mu}} \ln X_{\tilde{\mu}}
\: - \: t_a \right) \: + \: \sum_i \exp\left( - Y^i \right)
\: + \: \sum_{\tilde{\mu}} X_{\tilde{\mu}},
\end{equation}
where $\rho_i$ is a weight vector, and $\alpha_{\tilde{\mu}}$ is
a root vector.  In brief, the idea of the nonabelian mirror is that it
is abelian mirror symmetry in the Cartan torus, at a generic point on the
Coulomb branch.

In principle, both these mirrors have the property that correlation functions
in the original A-twisted GLSM are the same as correlation functions in the
B-twisted Landau-Ginzburg mirror.  We can derive a mirror map for operators
from the critical loci of the superpotential~(\ref{eq:W:nonabelian:mirror}).
From $\partial W/\partial X_{\tilde{\mu}} = 0$, we get
\begin{equation}
X_{\tilde{\mu}} \: = \: \sum_{a=1}^r \sigma_a \alpha^a_{\tilde{\mu}},
\end{equation}
and from $\partial W / \partial Y^i = 0$, we get
\begin{equation}
\exp\left( - Y^i \right) \: = \: \sum_{a=1}^r \sigma_a \rho_i^a.
\end{equation}
In both of these critical locus equations, the left-hand-side can be
interpreted in the B-twisted mirror, and the right-hand-side can be
interpreted in the original A-twisted GLSM.

Now, let us work through two examples. 
As before, we begin with the GLSM for ${\mathbb P}^n$.
The mirror \cite{Hori:2000kt} is a Landau-Ginzburg model with superpotential
\begin{equation}
W \: = \: \sigma\left( \sum_i Y^i - t \right) \: + \:
\exp\left( - Y^1 \right) \: + \: \cdots \: + \:
\exp\left( - Y^{n+1} \right).
\end{equation}
We can integrate out $\sigma$ and $Y^{n+1}$ to write
\begin{equation}
W \: = \: \exp\left( - Y^1 \right) \: + \: \cdots \: + \:
\exp\left( - Y^n \right) \: + \: q \exp\left( Y^1 + \cdots + Y^n \right),
\end{equation}
where $q = \exp(-t)$.
The critical locus is computed from
\begin{equation}
\frac{\partial W}{\partial Y^i} \: = \: 
- \exp\left( - Y^i \right) + q \exp\left( Y^1 + \cdots + Y^n \right) \: = \: 0,
\end{equation}
which implies
\begin{equation}
\exp\left( - Y^i \right) \: = \: q \prod_j \exp\left( + Y^j \right),
\end{equation}
so if we define $X = \exp(-Y^i)$, then
\begin{equation}
X^{n+1} \: = \: q,
\end{equation}
the ring relation in the quantum cohomology ring for ${\mathbb P}^n$.

Next, we turn to the Grassmannian $G(k,n)$.  
Here, the mirror \cite{Gu:2018fpm} is the
$S_k$ orbifold of a Landau-Ginzburg model with superpotential
\begin{eqnarray}
W & = &
\sum_{a=1}^k \sigma_a \left( \sum_{ib} \rho_{ib}^a Y^{ib} \: - \:
\sum_{\mu \neq \nu} \alpha^a_{\mu \nu} \ln X_{\mu \nu} \: - \: t
\right) \: + \: \sum_{ia}\exp\left( - Y^{ia} \right) \: + \:
\sum_{\mu \neq \nu} X_{\mu \nu},
\nonumber \\
& = &
\sum_{a=1}^k \sigma_a \left( \sum_a Y^{ia} \: + \:
\sum_{\nu \neq a} \left( \frac{ X_{a \nu} }{ X_{\nu a} } \right)
 \: - \: t \right)
\: + \: \sum_{ia} \exp\left( - Y^{ia} \right) \: + \:
\sum_{\mu \neq \nu} X_{\mu \nu},
\end{eqnarray}
where
\begin{equation}
\rho^a_{ib} \: = \: \delta^a_b, \: \: \:
\alpha^a_{\mu \nu} \: = \: - \delta^a_{\mu} + \delta^a_{\nu}.
\end{equation}
We integrate out $\sigma_a$, $Y^{na}$ to obtain
\begin{equation}
W \: = \: \sum_{i=1}^{n-1} \sum_{a=1}^k \exp\left( - Y^{ia} \right) \: + \:
\sum_{\mu \neq \nu} X_{\mu \nu} \: + \: \sum_{a=1}^k \Pi_a,
\end{equation}
where
\begin{equation}
\Pi_a \: = \: \exp\left( - Y^{na} \right) \: = \:
q \left( \prod_{i=1}^{n-1} \exp\left( + Y^{ia} \right) \right)
\left( \prod_{\nu \neq a} \frac{ X_{a \nu} }{ X_{\nu a} } \right).
\end{equation}

Next, we compute the critical locus.
From
\begin{equation}
\frac{\partial W}{\partial Y^{ia}} \: = \: - \exp\left( - Y^{ia} \right)
\: + \: \Pi_a \: = \: 0,
\end{equation}
we find
\begin{equation}
\exp\left( - Y^{ia} \right) \: = \: \Pi_a
\end{equation}
for all $i$.  Similarly, from
\begin{equation}
\frac{\partial W}{\partial X_{\mu \nu} } \: = \: 1 \: + \:
\frac{\Pi_{\mu} - \Pi_{\nu}}{X_{\mu \nu}} \: = \: 0,
\end{equation}
we find
\begin{equation}
X_{\mu \nu} \: = \: - \Pi_{\mu} + \Pi_{\nu},
\end{equation}
hence
\begin{equation}  \label{eq:gkn:mirror:critlocus}
\prod_{\nu \neq a} \frac{ X_{a \nu} }{ X_{\nu a} } \: = \: (-)^{k-1},
\: \: \:
\left( \Pi_a \right)^n \: = \: (-)^{k-1} q.
\end{equation}
The operator mirror map is
\begin{eqnarray}
\exp\left( - Y^{ia} \right) \: = \: \Pi_a & \leftrightarrow & \sigma_a,
\\
X_{\mu \nu} & \leftrightarrow & - \sigma_{\mu} + \sigma_{\nu},
\end{eqnarray}
so the critical locus equation~(\ref{eq:gkn:mirror:critlocus})
recovers the expression for the ring relation in the quantum cohomology
ring of $G(k,n)$ described earlier; in other words,
\begin{equation}
\left( \Pi_a \right)^n \: = \: (-)^{k-1} q
\end{equation}
becomes
\begin{equation}
\left( \sigma_a \right)^n \: = \: (-)^{k-1} q.
\end{equation}

Also, poles in the superpotential at $X_{\mu \nu} = 0$ correspond to the
excluded locus
\begin{equation}
\sigma_{\mu} \neq \sigma_{\nu}
\end{equation}
for $\mu \neq \nu$.

On a related matter, there was a talk at the meeting on nonabelian T-duality
by N.~Cabo Bizet.

In passing, we would also like to mention two other important topics,
which lack of space prevents us from describing in more detail:
\begin{itemize}
\item {\bf Supersymmetric localization.}

Supersymmetric localization was first applied to two-dimensional
GLSMs in, to our knowledge \cite{Benini:2012ui,Doroud:2012xw},
and was quickly applied to give alternative physical computations
of Gromov-Witten invariants \cite{Jockers:2012dk},
elliptic genera \cite{Benini:2013nda,Benini:2013xpa},
and Gamma classes \cite{Halverson:2013qca,libgober,iritani1,iritani2,kkp}.
These are important contributions, which we wanted to acknowledge,
but lack of space prevents us from going into any detail.

\item {\bf D-branes in GLSMs.}

GLSMs on open strings were explored in detail in
\cite{Herbst:2008jq}, which described e.g.~the grade restriction rule.
There is not space in this overview to explain any details,
but this was discussed at the meeting in talks by I.~Brunner, K.~Hori, J.~Guo,
and K.~Aleshkin.
\end{itemize}

\section{Quantum sheaf cohomology}

So far we have reviewed progress in GLSMs for two-dimensional theories
with (2,2) supersymmetry.  There also exist GLSMs for two-dimensional
theories with (0,2) supersymmetry 
\cite{Witten:1993yc,Distler:1993mk,Distler:1995mi}.  
Briefly, in geometric phases,
these describe a space $X$, along with a holomorphic vector bundle
${\cal E} \rightarrow X$, obeying the constraint
\begin{equation}
{\rm ch}_2({\cal E}) \: = \: {\rm ch}_2(TX).
\end{equation}
These theories admit analogues \cite{Katz:2004nn,Sharpe:2006qd}
of the A, B model topological
twists \cite{Witten:1991zz}:
\begin{itemize}
\item The analogue of the A twist, known as the A/2 model, exists when
$\det {\cal E}^* \cong K_X$, and has operators corresponding to elements
of $H^{\bullet}(X, \wedge^{\bullet} {\cal E}^*)$.
\item The analogue of the B twist, known as the B/2 model, exists when
$\det {\cal E} \cong K_X$, and has operators corresponding to elements
of $H^{\bullet}(X, \wedge^{\bullet} {\cal E})$.
\end{itemize}
These theories have (0,2) supersymmetry and reduce to the ordinary
A, B models in the special case that ${\cal E} = TX$.

The OPEs of local operators in these theories also describe generalizations
of quantum cohomology, known as quantum sheaf cohomology,
see e.g.~\cite{Katz:2004nn,Sharpe:2006qd,Adams:2005tc}.
We outline the details here.

First, recall that local operators in the ordinary A model with target
space $X$ correspond to elements of $H^{\bullet,\bullet}(X) = 
H^{\bullet}(X, \wedge^{\bullet} T^*X)$, and correlation functions
are computed mathematically by intersection theory on a moduli space of
curves.

Quantum sheaf cohomology \cite{Katz:2004nn,Sharpe:2006qd,Adams:2005tc}
arises from an A/2-twisted theory, with target space
$X$ and bundle ${\cal E}$  Local operators correspond to elements of
$H^{\bullet}(X, \wedge^{\bullet} {\cal E}^*)$.  These have a classical
product 
\begin{equation}
H^{\bullet}(X, \wedge^{\bullet} {\cal E}^*)
\times
H^{\bullet}(X, \wedge^{\bullet} {\cal E}^*) 
\: \longrightarrow \:
H^{\bullet + \bullet}(X, \wedge^{\bullet+\bullet} {\cal E}^*).
\end{equation}
Correlation functions are computed by sheaf cohomology on a moduli space
of curves, and the resulting local operator OPEs describe a 
deformation of the classical product structure above.
This reduces to ordinary quantum cohomology in the special case that
${\cal E} = TX$.

To be concrete, we outline a family of examples on ${\mathbb P}^1 \times
{\mathbb P}^1$.  First, recall the ordinary quantum cohomology
ring is
\begin{equation}  \label{eq:qh:p1p1}
{\mathbb C}[x,y] / \left( x^2 - q_1, y^2 - q_2 \right).
\end{equation}
Now, to define quantum sheaf cohomology, we must define a suitable
bundle ${\cal E}$.  Take ${\cal E}$ to be a deformation of the tangent
bundle, described as the cokernel
\begin{equation}
0 \: \longrightarrow \: {\cal O}^2 \: \stackrel{*}{\longrightarrow} \:
{\cal O}(1,0)^2 \oplus {\cal O}(0,1)^2 \: \longrightarrow \:
{\cal E} \: \longrightarrow \: 0,
\end{equation}
where
\begin{equation}
* \: = \: \left[ \begin{array}{cc}
A w & B w \\ C z & D z \end{array} \right],
\end{equation}
for $A$, $B$, $C$, $D$ constant $2 \times 2$ matrices (subject to
obvious nondegeneracy constraints) and $w$, $z$
column vectors of homogeneous coordinates on either ${\mathbb P}^1$
factor.  Then, the quantum sheaf cohomology ring of
${\mathbb P}^1 \times {\mathbb P}^1, {\cal E})$ is given by
\cite{McOrist:2007kp,McOrist:2008ji,Donagi:2011uz,Donagi:2011va}
\begin{equation} \label{eq:qsc:p1p1}
{\mathbb C}[x,y] / \left( \det(A x + B y) - q_1,
\det(C x + D y) - q_2 \right).
\end{equation}
When for example $A = D = I, B = C = 0$, then ${\cal E} = TX$
and the quantum sheaf cohomology ring~(\ref{eq:qsc:p1p1}) reduces
to the ordinary quantum cohomology ring~(\ref{eq:qh:p1p1}).

One way to compute quantum sheaf cohomology, for Fano spaces, is using
GLSMs and Coulomb branches \cite{McOrist:2007kp,McOrist:2008ji}.
The basic idea is the same as in (2,2) supersymmetry: under RG flow,
the GLSM flows onto a Coulomb branch where the OPE ring relations can
be computed as the critical locus of a twisted one-loop effective
superpotential.

In abelian cases, the resulting twisted superpotential is of the form
\begin{equation}
\tilde{W}(\sigma) \: = \:
\sum_a \Upsilon_a \ln\left( q_a^{-1} \prod_i (\det M_i(\sigma) )^{Q_i^a} \right),
\end{equation}
where $M_i(\sigma_a)$ are matrices encoding tangent bundle deformations,
and $\Upsilon_a$ is a (0,2) Fermi superfield (part of the (2,2) vector
multiplet).
The critical locus equations are
\begin{equation}
\frac{\partial \tilde{W}}{\partial \Upsilon_a} \: = \: 0
\end{equation}
which imply 
\begin{equation}  \label{eq:qsc:genlform}
\prod_i \left( \det M_i(\sigma) \right)^{Q_i^a} \: = \: q_a.
\end{equation}

We have already discussed ${\mathbb P}^1 \times {\mathbb P}^1$
examples, for which the quantum sheaf cohomology ring relations are
\begin{equation}
\det(A x + B y) \: = \: q_1, \: \: \:
\det(C x + D y) \: = \: q_2,
\end{equation}
the same form as~(\ref{eq:qsc:genlform}).

Another example is the Grassmannian $G(k,n)$.  Let ${\cal E}$ be a
deformation of the tangent bundle, defined by the cokernel
\begin{equation}
0 \: \longrightarrow \: S^* \otimes S \: \stackrel{*}{\longrightarrow} \:
{\mathbb C}^n \otimes S \: \longrightarrow \: {\cal E} \:
\longrightarrow \: 0,
\end{equation}
where
\begin{equation}
*: \: \omega_a^b \: \mapsto \: A^i_j \omega_a^b \phi^j_b \: + \:
\omega_b^b B^i_j \phi^j_a.
\end{equation}

Then, the quantum sheaf cohomology ring relations are
\cite{Guo:2015caf,Guo:2016suk}
\begin{equation}
\det(A \sigma_a + B {\rm Tr}\, \sigma) \: = \: (-)^{k-1} q,
\end{equation}
which for ${\cal E} = TX$ reduce to
\begin{equation}
\left( \sigma_a \right)^n \: = \: (-)^{k-1} q,
\end{equation}
which defines the ring relation of the ordinary quantum cohomology
ring of $G(k,n)$, as discussed previously.

Quantum sheaf cohomology is now known for
\begin{itemize}
\item Fano toric varieties \cite{McOrist:2007kp,McOrist:2008ji,Donagi:2011uz,Donagi:2011va},
\item Grassmannians \cite{Guo:2015caf,Guo:2016suk},
\item flag manifolds \cite{Guo:2018iyr},
\end{itemize}
all with ${\cal E}$ given by a deformation of the tangent bundle.
(Sheaf cohomology on toric complete
intersections has also been discussd \cite{Lyu:2024qej}.)
More general cases are open questions.

There is also a notion of mirror symmetry for (0,2) supersymmetric theories,
known as (0,2) mirror symmetry.  Just as the original form of
mirror symmetry relates pairs of Calabi-Yau's $X$, $Y$,
(0,2) mirror symmetry relates pairs $(X, {\cal E})$, 
$(Y, {\cal F})$, where $X$, $Y$ are Calabi-Yau (not necessarily
mirror in the ordinary sense) and ${\cal E} \rightarrow X$,
${\cal F} \rightarrow Y$ are holomorphic bundles such that
\begin{equation}
{\rm ch}_2({\cal E}) \: = \: {\rm ch}_2(TX),
\: \: \:
{\rm ch}_2({\cal F}) \: = \: {\rm ch}_2(TY).
\end{equation}
The twisted theories are close related:
\begin{eqnarray}
\mbox{A/2 on } (X, {\cal E}) & = & \mbox{B/2 on } (Y, {\cal F}),
\\
H^{\bullet}(X, \wedge^{\bullet} {\cal E}^*) & = &
H^{\bullet}(Y, \wedge^{\bullet} {\cal F}),
\end{eqnarray}
which for ${\cal E} = TX$, ${\cal F} = TY$, reduces to the standard
relation between the ordinary A, B models on mirrors, and 
the standard relation between Hodge diamonds.

(0,2) mirror symmetry has been studied for many years.
For example, numerical evidence was described in \cite{Blumenhagen:1996vu}.
There
are (limited) proposals for mirror constructions,
see e.g.~\cite{Blumenhagen:1996vu,Blumenhagen:1996tv,Adams:2003zy,Melnikov:2010sa,Gu:2019byn}.

For (0,2) GLSMs describing Fano spaces, (limited) proposals exist for
(0,2) mirrors as (0,2) Landau-Ginzburg models.  Consider for example
the case of ${\mathbb P}^1 \times {\mathbb P}^1$, with bundle ${\cal E}$
given as the cokernel
\begin{equation}
0 \: \longrightarrow \: {\cal O}^2 \: \stackrel{*}{\longrightarrow} \:
{\cal O}(1,0)^2 \oplus {\cal O}(0,1)^2 \: \longrightarrow \: {\cal E}
\: \longrightarrow \: 0,
\end{equation}
where 
\begin{equation}
* \: = \: \left[ \begin{array}{cc}
A w & B w \\ C z & D z \end{array} \right],
\end{equation}
as before.  If we restrict to diagonal matrices $A$, $B$, $C$, $D$,
then a mirror (0,2) Landau-Ginzburg model is defined by
\begin{eqnarray}
W & = & \Upsilon\left( Y_0 + Y_1 - t_1 \right) \: + \:
\tilde{\Upsilon} \left( \tilde{Y}_0 + \tilde{Y}_1 - t_2 \right)
 \\
& & \hspace*{0.25in}
\: + \:
\sum_{i=1}^1 F_i \left( E_i(\sigma, \tilde{\sigma}) - \exp(-Y_i) \right)
\: + \:
\sum_{j=0}^1 \tilde{F}_j \left( \tilde{E}_j(\sigma, \tilde{\sigma}) - 
\exp(-\tilde{Y}_j) \right),
\nonumber
\end{eqnarray}
where
\begin{equation}
E_i(\sigma, \tilde{\sigma}) \: = \: a_i \sigma + b_i \tilde{\sigma},
\: \: \:
\tilde{E}_j(\sigma, \tilde{\sigma}) \: = \: c_i \sigma + d_i \tilde{\sigma},
\end{equation}
\begin{equation}
A \: = \: {\rm diag}(a_0,a_1), \: \: \:
B \: = \: {\rm diag}(b_0,b_1), \: \: \:
C \: = \: {\rm diag}(c_0,c_1), \: \: \:
D \: = \: {\rm diag}(d_0,d_1),
\end{equation}
$\Upsilon_i$, $F_i$, $\tilde{\Upsilon}_j$, $\tilde{F}_j$ are (0,2)
Fermi superfields, parts of (2,2) $\sigma$ and $Y$ multiplets.

There were several talks at this meeting on various aspets of 2d (0,2) theories,
including talks of S.~Gukov, M.~Litvinov, and S.~Franco.

In passing, we would also like to mention two other important topics,
which lack of space prevents us from describing in more detail:
\begin{itemize}
\item {\bf Triality.} Triality is a property
of (0,2) supersymmetric
theories, first discussed in \cite{Gadde:2013lxa}.
This is an IR duality relating triples of theories.
They have the following prototypical form.
Briefly, a (0,2) theory describing the Grassmannian $G(k,n)$ with
bundle
\begin{equation}
S^{\oplus N} \oplus (Q^*)^{2k+N-n} \oplus (\det S^*)^{\oplus 2}
\end{equation}
(for $S$ the universal subbundle and $Q$ the universal quotient bundle)
is IR equivalent to a (0,2) theory describing the Grassmannian $G(n-k,N)$
with bundle
\begin{equation}
S^{\oplus 2k + N - n} \oplus (Q^*)^n \oplus (\det S^*)^{\oplus 2},
\end{equation}
and is also IR equivalent to a (0,2) theory describing the Grassmannian
$G*N-n+k, 2k+N-n)$ with bundle
\begin{equation}
S^{\oplus n} \oplus (Q^*)^N \oplus (\det S^*)^{\oplus 2},
\end{equation}
for $k$, $n$, $N$ satisfying certain inequalities, which simultaneously
guarantee both that the geometric description is sensible, and that
supersymmetry is unbroken.

Triality was discussed further in S.~Franco's talk.

\item {\bf GLSMs with $H$ flux.}
These have a long history
\cite{Adams:2006kb,Adams:2009av,Adams:2009zg,Adams:2012sh,Quigley:2011pv,Melnikov:2012nm,Caldeira:2018ynv}, and are often used to describe,
for example, non-K\"ahler heterotic compactifications.
The details are well beyond the scope of this short overview, but certainly
deserve to be mentioned.
\end{itemize}

\section{Quantum K theory}

Just as two-dimensional GLSMs can sometimes be used to
compute quantum cohomology,
it has been noted \cite{Bullimore:2014awa,Jockers:2018sfl,Jockers:2019wjh,Ueda:2019qhg}
that three-dimensional GLSMs can sometimes be used to compute quantum K theory.
Furthermore, analogous to other examples in this survey, 
in many cases quantum K theory
can be computed using Coulomb branch techniques.

The basic idea of the physical realization of quantum K theory
is as follows (see for example \cite{Bullimore:2014awa,Jockers:2018sfl,Jockers:2019wjh,Ueda:2019qhg}).  
Consider a GLSM in three dimensions,
on a three-manifold of the form $S^1 \times \Sigma_2$, where $\Sigma_2$ is
a Riemann surface.  Quantum K theory arises as OPEs of Wilson lines wrapped
on the $S^1$, moving parallel to one another along the base $\Sigma_2$.

To compute those OPEs, one does a Kaluza-Klein reduction \cite{Nekrasov:2009uh} 
along the $S^1$.
One gets an effective low-energy two-dimensional theory (along $\Sigma_2$),
with an infinite tower of fields.
Regularizing the sum of their contributions to the two-dimensional
twisted one-loop effective superpotential has the effect of changing
the ordinary log contributions to dilogarithms Li$_2$.

The Wilson line OPE relations are the critical loci of the two-dimensional
twisted superpotential \cite{Closset:2016arn,Closset:2018ghr,Closset:2019hyt,Ueda:2019qhg}.

Let us work through a simple example.
Consider a three-dimensional GLSM for ${\mathbb P}^n$,
meaning a $U(1)$ gauge theory with $n+1$ chiral superfields of charge $+1$.
The twisted one-loop effective superpotential for the two-dimensional
theory, obtained after regularizing the sum of Kaluza-Klein states,
and for the pertinent Chern-Simons level, is of the form
\begin{equation}
\tilde{W} \: = \: \left( \ln q \right) \left( \ln x \right) \: + \:
\sum_{i=1}^{n+1} {\rm Li}_2(x),
\end{equation}
where $x = \exp(2 \pi i R \sigma)$ for $R$ the radius of the $S^1$,
and $\sigma$ the scalar of the two-dimensional vector multiplet.
The critical locus of this superpotential is 
\begin{equation}  \label{eq:qk:pn}
\left( 1 - x \right)^{n+1} \: = \: q.
\end{equation}
This is precisely the quantum K theory ring relation for 
${\mathbb P}^n$, where we identify $x$ with $S = {\cal O}(-1)$, the
tautological line bundle.  (Classically, in K theory,
$1 - S = {\cal O}_D$ for $D$ a hyperplane divisor, and the
$(n+1)$-fold self-intersection of a divisor on an $n$-dimensional space 
vanishes.)
(Superpotentials for more general cases has also been discussed 
\cite{Gu:2020zpg,Closset:2016arn,Nekrasov:2009uh}.)

We can relate the quantum K theory ring relation to the quantum cohomology
ring relation, in the limit that $R \rightarrow 0$.  To that end, in that
limit, expand
\begin{equation}
x = \exp(2 \pi i R \sigma) \: \mapsto \: 1 + 2 \pi i R \sigma,
\: \: \:
q = R^{d+1} q_{2d},
\end{equation}
and it is straightforward to see that the ring relation~(\ref{eq:qk:pn})
reduces to
\begin{equation}
\sigma^{n+1} \propto q_{2d},
\end{equation}
which is the standard quantum cohomology ring relation for
${\mathbb P}^n$.

For another example, we turn to the Grassmannian $G(k,n)$.
For the pertinent Chern-Simons level, the twisted one-loop
effective superpotential, after regularizing the sum over 
Kaluza-Klein modes, is given by
\begin{equation}
\tilde{W} \: = \:
\frac{k}{2} \sum_{a=1}^k \left( \ln x_a \right)^2 \: - \:
\frac{1}{2} \left( \sum_{a=1}^k \ln x_a \right)^2 \: + \:
\left( \ln (-)^{k-1} q \right) \sum_{a=1}^k \ln x_a
\: + \: n \sum_{a=1}^k {\rm Li}_2(x_a),
\end{equation}
where $x_a = \exp(2 \pi i R \sigma_a)$, for $R$ the radius of the
$S^1$, and $\sigma_a$ the vev of the scalar in the two-dimensional
vector multiplet on the Coulomb present.
(Also present, though not written explicitly, are the
Weyl-group ($S_k$) orbifold, and the excluded locus
$\sigma_a \neq \sigma_b$.)

The critical locus of this superpotential is
\begin{equation}
\left( 1 - x_a \right)^n \left( \prod_{b=1}^k x_b \right)
\: = \:
(-)^{k-1} q (x_a)^k.
\end{equation}
This equation can be symmetrized as before using Vieta, to obtain
\begin{equation}  \label{eq:qk:gkn:basic}
\sum_{r=0}^{n-i} e_{\ell - r}(x) e_r(\overline{x}) \: = \:
\left( \begin{array}{c} n \\ \ell \end{array} \right)
+ q e_{n-k}(\overline{x}) \delta_{\ell, n-k}.
\end{equation}

One can show \cite{Gu:2023tcv} 
that the symmetric polynomials in the $\overline{x}$
are interpreted as coupling to 
\begin{equation}
e_{\ell}(\overline{x}) \: = \:
\left\{ \begin{array}{cl}
\wedge^{\ell} ( {\mathbb C}^n/S) & \ell < n-k,
\\
(1-q)^{-1} \wedge^{\ell} ( {\mathbb C}^n/S) & \ell = n-k,
\end{array} \right.
\end{equation}
so the ring relations~(\ref{eq:qk:gkn:basic}) become
\begin{eqnarray}
\lefteqn{
\sum_{r=0}^{n-k-1} \wedge^{\ell-r}(S) \star \wedge^r( {\mathbb C}^n/S ) 
\: + \:
\frac{1}{1-q} \wedge^{\ell - (n-k)} S \star \det( {\mathbb C}^n/S)
} \nonumber \\
 & \hspace*{1in}  = &
\wedge^{\ell} {\mathbb C}^n \: + \:
\frac{1}{1-q} \det( {\mathbb C}^n/S) \, \delta_{\ell, n-k},
\end{eqnarray}
or after simplification,
\begin{equation}
\lambda_y(S) \star \lambda_y( {\mathbb C}^n/S) \: = \:
\lambda_y( {\mathbb C}^n) \: - \:
y^{n-k} \frac{q}{1-q} \det( {\mathbb C}^n/S) \star \left(
\lambda_y(S) - 1 \right),
\end{equation}
where $\star$ denotes the quantum product, and
\begin{equation}
\lambda_y({\cal E}) \: = \:
1 + y {\cal E} + y^2 \wedge^2 {\cal E} + y^3 \wedge^3 {\cal E}
+ \cdots.
\end{equation}
This is a presentation\footnote{
To be clear, the quantum K theory ring of $G(k,n)$ has been studied from
a variety of perspectives in both the math and physics communities;
see for example \cite{gl} for an early mathematics reference, and
see for example \cite{Ueda:2019qhg} for an early physics reference.
} of the quantum K theory ring of the Grassmannian
$G(k,n)$ \cite{Gu:2020zpg,Gu:2022yvj}.

There exists an analogous presentation of the quantum K theory ring
of partial flag manifolds, of the form \cite{Gu:2023tcv,Gu:2023fpw}
\begin{equation}
\lambda_y(S_i) \star \lambda_y( S_{i+1}/S_i) \: = \:
\lambda_y(S_{i+1}) \: - \:
y^{k_{i+1} - k_i} \frac{q_i}{1-q_i} \det( S_{i+1}/S_i) \star
\left( \lambda_y(S_i) - \lambda_y(S_{i-1}) \right),
\end{equation}
where $S_i$ is a universal subbundle of rank $k_i$.
Weihong Xu's talk at this meeting described this in greater detail.

In this discussion, we have mostly glossed over the role of Chern-Simons
levels.  The three-dimensional supersymmetric theory can certainly have
Chern-Simons terms, and their levels modify the low-energy twisted one-loop
effective superpotential $\tilde{W}$.  We have chosen Chern-Simons levels
in the expressions above to match quantum K theory results, 
corresponding to $U(1)_{-1/2}$ quantization of the chirals
\cite[section 2.2]{Closset:2019hyt},
but one can
also choose other values for the levels.  It is believed that other
choices correspond to the mathematical notion of levels
discussed in \cite{rz}, but a detailed dictionary is not known for all
cases.

We have also glossed over Wilson line OPEs for more general cases,
not necessarily associated with quantum K theory.
These have been extensively studied in the literature,
see e.g.~\cite{Closset:2018ghr,Closset:2019hyt} and references therein.

Earlier we discussed the role of ordinary mirror symmetry and (0,2)
mirror symmetry in computing e.g.~quantum cohomology.
Similarly, there is a notion of mirror symmetry in three-dimensional
gauge theories, see for example 
\cite{Intriligator:1996ex,Dorey:1999rb,Aganagic:2001uw,Aharony:2017adm,rwz,Comi:2022aqo}.
The details are, unfortunately, beyond the scope of this short survey.

Others at this meeting who spoke on various aspects of quantum K theory
included P.~Koroteev, Y.~P.~Lee, and W.~Xu, 
and related work in three-dimensional
gauged linear sigma models was discussed by C.~Closset, H.~Jockers, 
and M.~Litvinov.
There were also discussions of related notions in integrable systems
in the talks of P.~Koroteev and W.~Gu.

\section{Conclusions}

In this overview we have surveyed a few relatively recent
developments in the physics of
gauged linear sigma models.

One question for the future is whether 
quantum K theory and quantum sheaf cohomology can be linked?
The boundary of a three-dimensional $N=2$ theory is a two-dimensional
(0,2) supersymmetric theory \cite{Yoshida:2014ssa,Dimofte:2017tpi,Dimofte:2019buf,Costello:2020ndc}.
One could imagine moving bulk operators to the boundary and using
the bulk/boundary correspondence to describe quantum sheaf cohomology
(of the two-dimensional (0,2) boundary) as a module over quantum K theory
(of the three-dimensional $N=2$ bulk).  However, one issue is that the
bulk operators are Wilson lines, not local operators, unlike the boundary;
moving those bulk operators to the boundary would yield Wilson lines in
the two-dimensional (0,2) supersymmetric boundary.  To implement this
program would require a mathematical interpretation of two-dimensional
(0,2) Wilson lines in terms of (presumably descendants in) quantum sheaf
cohomology.

One direction we have not surveyed are the newer mathematically-rigorous
approaches to GLSMs \cite{fjr1,fjr2,fjr3,fjr4}.  
These are extremely interesting, but there is not enough space here to
survey them.
Those constructions were described in talks by
H.~Fan, E.~Segal, C.~C.~Melissa Liu, and D.~Favero.

\section*{Acknowledgments}

E.S.~would like to thank the Simons Center for hosting this conference,
{\it GLSMs@30}.
E.S.~was partially supported by NSF grant PHY-2310588.



\end{document}